\documentclass[pra,reprint,nofootinbib,showpacs,superscriptaddress,twocolumn]{revtex4-1}

\usepackage{graphicx}
\usepackage{dcolumn}
\usepackage{bm}

\usepackage{mathtools}
\usepackage{color}
\usepackage{graphicx}
\usepackage{dcolumn}
\usepackage{amsbsy}
\usepackage{amsfonts}
\usepackage{color,soul}
\usepackage{amssymb}
\usepackage{bm}

\newcommand{\be}{\begin{equation}}
\newcommand{\ee}{\end{equation}}
\newcommand{\bea}{\begin{eqnarray}}
\newcommand{\eea}{\end{eqnarray}}

\newcommand{\Tr}{{\rm Tr\,}}

\newcommand{\<}{\langle}
\renewcommand{\>}{\rangle}
\renewcommand{\Re}{{\rm Re\,}}
\renewcommand{\Im}{{\rm Im\,}}
\renewcommand{\vec}[1]{{\bf #1}}
\renewcommand{\phi}{\varphi}
\renewcommand{\epsilon}{\varepsilon}

\newcommand{\ve}{\varepsilon}

\begin{document}


\title{Hydrodynamic behavior of a non-interacting quantum particles\\ in  presence of dephasing}

\author{Oles Shtanko}
\affiliation{Physics Department, Massachusetts Institute of Technology, Cambridge, Massachusetts 02139, USA}
\author{Seth Lloyd}
\affiliation{Department of Mechanical Engineering, Massachusetts Institute of Technology, Cambridge, Massachusetts 02139, USA}

\begin{abstract}
In solids and organic materials, environment-induced dephasing of 
particles and long-lived 
excitations leads to the crossover in their transport properties between 
quantum wave-like propagation and classical diffusive motion.  In this work, 
we demonstrate that dynamics of single carriers in this intermediate
crossover regime can exhibit distinct signatures such as the formation 
of vortices and viscous flow, the phenomena typically considered as 
manifestations of hydrodynamic transport.   We explain this effect
by modeling suppressed quantum interference of carriers, and we show
that the resulting dynamics resembles the linearized Navier-Stokes equations. 
Dephasing-assisted viscosity provides a potential alternative
explanation of the results of recent experiments exhibiting hydrodynamic
behavior in solids, and suggests experimental probes of how quantum
carriers couple to their environment.
\end{abstract}

\maketitle


The quantum nature of particles and long-lived excitations in solids leads 
to a number of non-trivial phenomena including but not limited to 
quantized transport  \cite{datta1997electronic,ando2012mesoscopic}, 
ballistic propagation of heat \cite{casimir1938note}, ergodicity breaking 
\cite{nandkishore2015many}, etc. Reliable observation of these phenomena, 
according to standard intuition, requires the system to be well isolated 
from the environment (including phonons). Indeed, the environment typically 
destroys quantum coherence in the system leading to the emergence of 
classical behavior \cite{breuer2002theory}. 
In some cases, however, the presence of environment is \textit{essential} to 
observe quantum effects.  Quantum phenomena that require environmental
interactions include, for example, topological states in non-Hermitian 
Hamiltonians \cite{rudner2009topological} and efficient exciton transport 
in organic molecular systems 
\cite{mohseni2008environment,rebentrost2009environment,plenio2008dephasing}. 
Such environment-assisted quantum phenomena typically 
arise in an intermediate regime where the dephasing rate is neither 
too strong nor too weak, a manifestation of what is
sometimes referred to as the Goldilocks effect \cite{lloyd2011quantum}.

In this work, we study a new quantum effect, environmenatlly induced quantum
viscosity (EQV), which arises from the competition 
between coherent quantum dynamics and the dephasing. We demonstrate that the probability current distribution of single quantum particle subject to dephasing may exhibit a dynamics similar to the behavior of the compressible viscous fluid. 
Instead of being induced by interaction with other particles, as in conventional hydrodynamics, environmentally induced quantum viscosity (EQV) arises from
the interplay between wavelike, ballistic behavior of quantum transport,
and a decohering environment. 
We apply our theory to 
different sample geometries and show that in a particular regime of dephasing rates, this fluid-like behavior leads to a vortex formation and Poiseuille flow in thin films,  signatures of hydrodynamic behavior detectable in modern transport experiments.

To examine the physical properties of the critical behavior, we model the dephasing process by a local random phase shift applied to single particle wavefunction at each infinitesimal time step \cite{haken1973exactly,altshuler1982effects}. In this model, the long-distance quantum correlations in the system are suppressed, and quantum interference vanishes at distances larger than the quasi-classical mean free path. This approximation makes it feasible to use a contracted system representation described in terms of local quantities such as probability density and current density. As we show below, the presence of short-lived quantum correlations, in this case, leads to corrections to the equations of motion. In the secular approximation, these corrections partially resemble the viscosity terms in the Navier-Stokes equation. The origin of these terms is related to a Huygens-Fresnel type interference from only neighboring points of wavefront resulting in a local equilibration in the group velocity of the single-particles wave function. As a result, the quantum corrections lead to an emergence of vorticity in that single-particle wave function, and to a non-local response to the external field, with the effect most evident when the mean free path is comparable with the size of the system. 

The regime we study in this work has several potential realizations in charge carriers in solids and acoustic phonons. In this case, the role of dephasing acting on the carriers is played by inelastic scattering processes breaking quasi-momentum conservation ($R$-processes) such as phonon scattering, disorder-assisted collisions, Umklapp scattering, etc. 
For weak dephasing,  the dynamics of carriers is approximately the ballistic propagation of plane waves or, alternatively, quantum random walks  \cite{aharonov1993quantum}. By contrast, high dephasing leads to a classical random walk dynamics such as the dynamics underlying charge diffusion.
When the parameters such as temperature and system size are varied, the system can make the transition into the regime intermediate between the ballistic and diffusive regimes, where dephasing-assisted viscosity effects play a significant role and lead to the quasi-hydrodynamic behavior.

At this point, it is instructive to compare the behavior of such dissipative non-interacting or indeed single particle quantum systems to the classical fluid dynamics of electrons \cite{0243-cao2011universal,0233-sommer2011universal}. The fluid-dynamic regime arises in strongly interacting systems if frequent mutual collisions of electrons preserve both total energy and the vector of the total momentum of quasiparticles ($N$-processes) \cite{1334-giuliani2005quantum,1143-kadanoff1963hydrodynamic}. These additional conservation laws allow classical electrical flow in clean systems to persist regardless of applied external potential. The latter property leads to the presence of non-local response and vorticity in certain system geometries \cite{0075-levitov2016electron,0013-falkovich2017linking}. Additionally, the hydrodynamic type of motion has several additional signatures detectable in experiments such as second sound \cite{0584-chester1963second,1665-joseph1989heat} and non-monotonic dependence of resistance on temperature (Gurzhi effect) \cite{0079-bass1990temperature,0117-gurzhi1968hydrodynamic}. First observed in low temperature solid helium \cite{0342-ackerman1966second}, hydrodynamic effects have seen recent confirmation in mesoscopic wires and films \cite{0123-DeJong1995,0096-moll2016evidence,0081-hehlen1995observation,0050-molenkamp1994electron}, graphene \cite{0159-crossno2016observation,0153-bandurin2016negative,0020-kumar2017superballistic,0011-molenkamp1994observation}, and systems of cold atoms \cite{0068-joseph2011observation}.  It is worth mentioning that fluid dynamics is also naturally successful in describing hot nuclear matter \cite{0103-jacak2012exploration,0091-Andreev2011}. As we show below, the methods developed to detect hydrodynamic transport should equally work to reveal the non-local effects in systems of non-interacting particles with the environment-induced dephasing.

\begin{figure}[t!]
\includegraphics[width=1.\columnwidth]{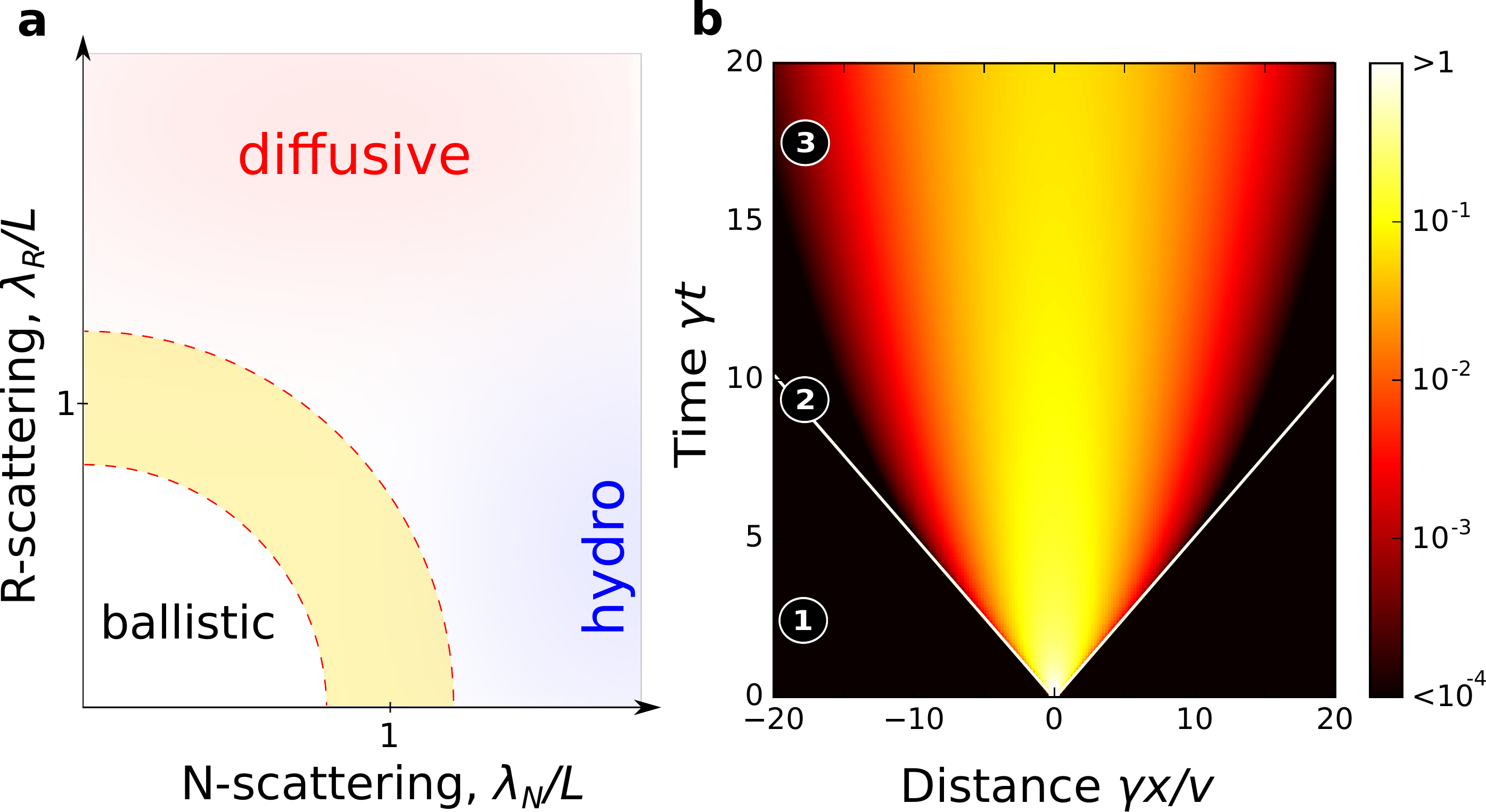}
\caption{\textbf{Quantum transport regimes.} \textbf{a,} Schematic diagram for quantum transport in the presence of the environment and particle collisions. The critical regime described by quasi-hydrodynamic transport is depicted by yellow area. \textbf{b,} The time evolution of the probability density of initially localized wavepacket described by Eq.$\,$\eqref{eq:master_equation}. During propagation, the wavepacket exhibits with time various regimes of propagation: ballistic (1), critical (2), and diffusive (3).}
\label{fig:transport_regimes}
\end{figure}

We show that the critical dephasing-assisted transport predicts results that are
surprisingly similar to fluid mechanics for non-local response in several types of sample geometry. 
In particular, we study the flow between two parallel surfaces with no-slip boundary conditions \cite{0011-jaggi1991electron} and 2D current flow between two narrow leads \cite{0075-levitov2016electron} where the non-local response is typically considered as a direct manifestation of collective transport. At the same time, these phenomena are essentially different. As mentioned before, the hydrodynamic equations are purely classical and require strong interactions between particles. By contrast, dephasing-assisted viscous transport relies on dephasing rtes in the intermediate Goldilocks regime, and and is characterized by quantum coherence preserved at a scale much smaller than the quasiclassical mean free path. The schematic diagram in Fig. \ref{fig:transport_regimes}a summarizes the conditions for ballistic, diffusive, hydrodynamic, and the critical regimes depending on the mean free path to system size ratios $\lambda_R/L$ and $\lambda_N/L$ for $R$- and $N$-processes respectively. 

\textbf{Model}. We consider the evolution of non-interacting carriers using the single particle Hamiltonian 
\be\label{eq:hamiltonian}
H(t) = \frac{\hbar v}{a}\sum^d_{i=1}\sum_{\vec r}\Bigl( |\vec r\>\<\vec r+a\vec e_i|+{\rm h.c.}\Bigl)+\sum_{\vec r}\xi_{\vec r}(t)|\vec r\>\<\vec r|,
\ee
where $a$ is the lattice spacing, $\{\vec e_i\}$  set of primitive vectors on $d$-dimensional square lattice, and $\xi_{\vec r}(t)$ is a fluctuating potential we approximate by unbiased zero-average gaussian random variable, $\bigl\<\xi_{\vec r_1}(t)\xi_{\vec r_2}(t')\bigl\> = \alpha\, \delta_{\vec r_1\vec r_2}\delta(t-t')$. The last term reflects the presence of $R$-processes in the system resulting in local random phase shifts in single particle wavefunction $\psi_{\vec r}(t) \to\exp(i\xi_{\vec r}(t)\delta t)\psi_{\vec r}(t)$ at each time step $\delta t$.
This method has been proposed first by Haken and Strobl \cite{haken1973exactly} in a context of Frenkel excitons dynamics in molecular crystals. The Haken-Strobl model is, however, equally well applicable to single carriers in metals and semiconductors,
and describes a system exhibiting a transition from the quantum ballistic to the diffusive regime.

The dynamics of the system can be characterized in terms of the averaged evolution over all possible statistical realizations of the fluctuating parameter $\xi_{\vec r}(t)$, namely
\be\label{eq:evolution}
\rho(t) = \Bigl\< U_t\rho(0)U_t^\dag\Bigl\>_\xi,\quad U_t = \mathcal T \exp\left(i\int_0^t dt' H(t') \right),
\ee
where $U_t$ is unitary time evolution operator, and ${\mathcal T\rm exp}$ denotes the time-ordered exponential. 

The process in Eq.$\,$\eqref{eq:evolution} belongs to the class of Markovian open quantum system dynamics \cite{breuer2002theory}. The most general case of this dynamics can be described in terms of a non-unitary Lindbladian master equation. In particular, the fluctuation-averaged dynamics in Eq.$\,$\eqref{eq:evolution} follows the quantum Boltzmann equation \cite{haken1973exactly,rebentrost2009environment}
\be\label{eq:master_equation}
\frac{\partial\rho}{\partial t} = -\frac{i}\hbar[H_0,\rho] + \gamma \sum_{\vec r} \left(\mathcal P_{\vec r} \rho \mathcal P_{\vec r} - \frac12 \mathcal P_{\vec r}\rho-\frac12\rho \mathcal P_{\vec r}\right),
\ee 
where $\gamma = \alpha/\hbar^2$ is the dephasing (scattering) rate, $\mathcal P_{\vec r}= |\vec r\>\<\vec r|$ are on-site projectors. The right-hand side of the equation is a sum of two distinct terms representing unitary dynamics and dissipative effects respectively. Considered separately, the unitary term describes the coherent propagation of particles over the lattice exhibiting quantum correlations between distant  lattice sites. Conversely, the dissipative term leads to exponential decay of any quantum correlation in the system. 
As we illustrate below, the competition between unitary dynamics and dephasing terms in the master equation is responsible for the crossover between quantum and classical propagation regimes.

\textbf{Dynamics}. To demonstrate the dephasing-assisted viscosity effects, we first derive the approximate equations of motion for local parameters such as probability density $P(\vec r,t)$ and the vector of current density $J_k(\vec r,t)$, where $k$ enumerates the space components. In the model Eq.$\,$\eqref{eq:master_equation}, one may connect the local parameters to the matrix elements of the density operator, $P(\vec r,t) = \rho(\vec r,\vec r)$ and $J_k(\vec r,t) = 2v\,\Im\rho(\vec r,\vec r +\vec e_k)$. The latter definition implies that the dynamics of the density following from Eq.$\,$\eqref{eq:master_equation} takes the form of the continuity equation
\be\label{eq:continuity}
\frac {\partial P}{\partial t}=- \nabla \cdot \vec J,
\ee
where we omit of the dependence on $\vec r$ and $t$, and we consider the continuous limit, $a\to0$ \footnote{In the continuous limit, we define the gradient operator for a function $f(\vec r)$ as $\nabla_k f(\vec r) \equiv \lim_{a\to0}(f(\vec r+a\vec e_k)-f(\vec r))/a$, while the integration over lattice space volume $\Omega$ is defined as a Riemann sum $\int_\Omega d\vec r \equiv a^d\sum_{\vec r\in \Omega}$.}.

In turn, the dynamics of the current density, as follows from Eq.$\,$\eqref{eq:master_equation}, is given by
\be\label{eq:current_evo}
\frac {\partial J_k}{\partial t} =-\gamma J_k-2v^2\nabla_k P  +  I_k,
\ee
where $ I_k\equiv  v\nabla_kF^+_{kk}+v\sum_{l\neq k}\nabla_l(F^+_{kl}-F^-_{kl})$, and the parameter $F^\pm_{kl} = 2\Re \rho(\vec r,\vec r +\vec e_k\pm\vec e_l)$ contains second diagonal elements of the density matrix (see \textit{SI: Part 1} for details). The first two terms in right-hand side of Eq.$\,$\eqref{eq:current_evo} describe conventional Ohmic dissipation leading to diffusive transport. In particular, the first term represents the decay of the current due to the presence of collisions violating momentum conservation. The second term supplies the current generated by the gradient of probability density. At the same time, the effects beyond the standard Ohmic regime originate in the term $I_k$ representing contributions from high order quantum correlations.

To complete the set of Eq.$\,$\eqref{eq:continuity} and Eq.$\,$\eqref{eq:current_evo}, one may express the correction $I_k$ as a linear functional of the current density
\be\label{eq:quantum_corrections}
 I_k[\vec J] =  \sum_{k'=1}^d\int_0^t dt'\int_V d\vec r' K_{kk'}(\vec r-\vec r',t-t') J_{k'}(\vec r',t'),
\ee
where $K_{kk'}(\vec r,t)$ is a memory function (see \textit{SI: Part 2} for details).

\begin{figure*}[t]
\includegraphics[width=1.8\columnwidth]{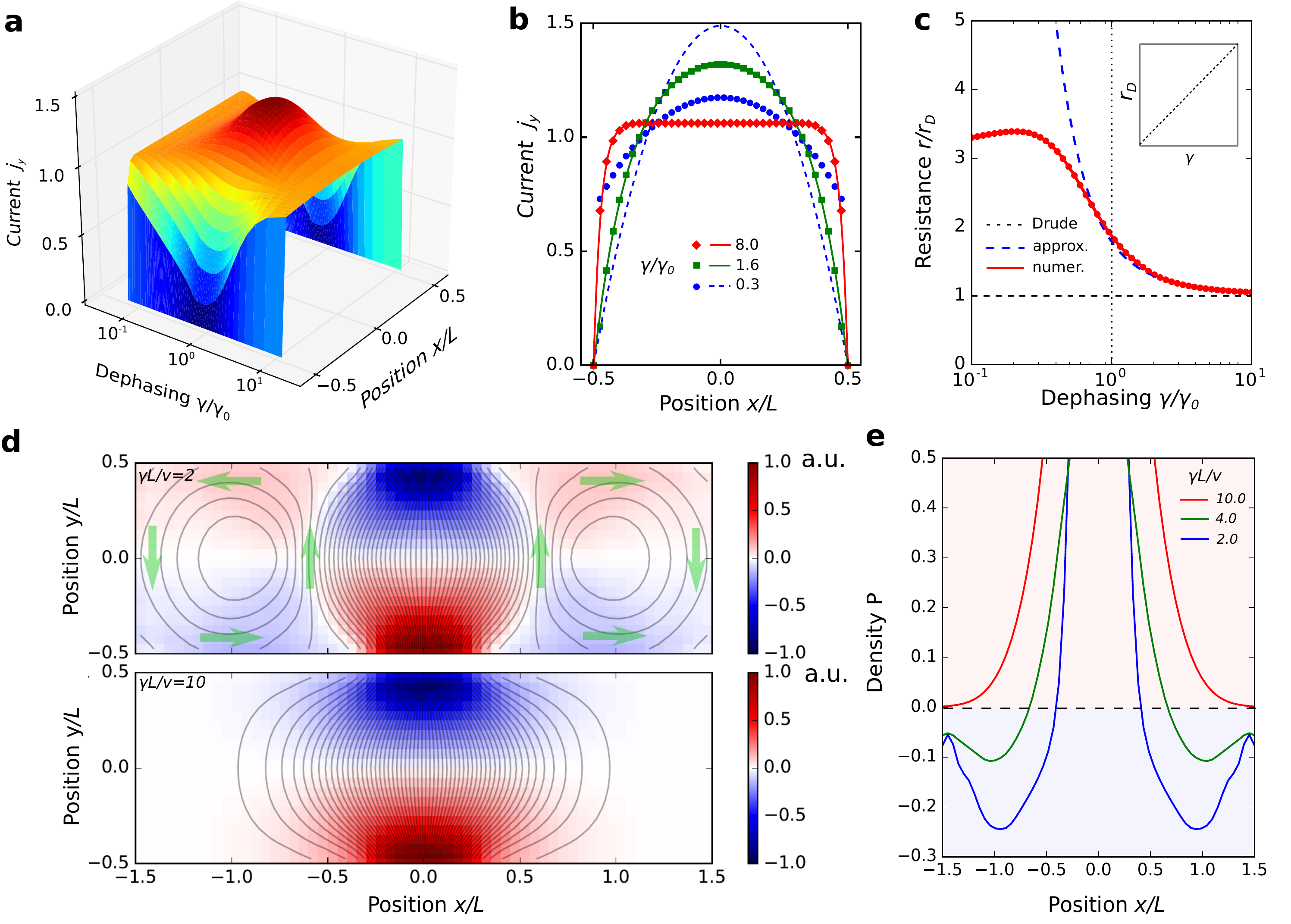}
\caption{\textbf{Effects of dephasing-assisted viscosity}. \textbf{a,} Evolution upon changing the dephasing rate $\gamma$ of current profile between two parallel surfaces with no-slip boundary condition. In the vicinity of the critical point $\gamma_0 = 2\pi v/L $ the profile has a Poiseulle-like form indicating the depahsing-induced viscous regime.
\textbf{b,} The current profile density in different regimes: nearly diffusive (red), critical (green), and nearly ballistic (blue).  Markers represent the numerical solution; the lines (including blue dashed line) represent the analytical approximation using Eq.$\,$ \eqref{eq:approx_dynamics}. \textbf{c,} Ratio of resistivity to the value predicted by the Drude law, $r_D = \gamma/2v^2$. Red curve represents numerical simulations, blue dashed curve is the analytical approximation in Eq.$\,$\eqref{eq:approx_dynamics}.
\textbf{d,} Vortex formation and non-local resistance in configuration with two leads supplying a source and a drain of particles. The panel shows the numerical result for steady-state density distribution for different values $\gamma L/v = 2$ and $\gamma L/v = 10$, the black curves represent the stream lines of the current. \textbf{e,} Dependence of density as a function of distance from the lead, as follows from numerics. For large $\gamma L/v$ (red curve) the sign of density is monotonic function, while for small $\gamma L/v$ (blue curve) the density changes sign. The latter is a direct signature of non-local response.}
\label{fig:figure_2}
\end{figure*}

In general, the derivation of the memory function is a difficult task. However, in limited cases, its expression can be found explicitly, for example, in one spatial dimension where the model is exactly solvable \cite{haken1973exactly,esposito2005emergence,vznidarivc2010exact}. As shown in $\textit{SI: Part 2}$, if initially at $t=0$ the system contains no quantum correlations (i.e. its density matrix is diagonal in coordinate basis), this function takes a compact form  in Fourier space with the wavevector $q$ and in Laplace space with frequency parameter $s$.
In this case, the kernel is given by the expression
$K(q,s) = \sqrt{(s+\gamma)^2/4+v^2q^2}-(s+\gamma)/2$. This exact solution
allows us to study analytically the different transport regimes in the model. 

To illustrate the effect of dissipation on the dynamics in 1D, it is instructive to derive the Green's function characterizing the spread of a wavepacket, $P(x,t) = \int_V dx' G(x-x',t)P(x',0)$. The Green's functions can be expressed as $G(q,s) = (s+2K(q,s))^{-1}$ or, in the space and time domain,
\be\label{eq:greens_function}
G(x,t) = \int_{-i\infty+c}^{i\infty+c} ds\int_{-\infty}^\infty\frac{dq}{2\pi}\frac{e^{iqx+st}}{\sqrt{(s+\gamma)^2+4v^2q^2}-\gamma}
\ee
where $c>0$, and we use the Mellin contour to derive the inverse Laplace transformation. The structure of the Green's function is shown in Fig.$\,$\ref{fig:transport_regimes}b. The panel demonstrates existence of two limits of propagation, ballistic and diffusive, separated by the critical crossover regime. These regimes are visibly distinct: the ballistic regime has a well defined lightcone (white lines) while the diffusive regime shows long suppressed tails for the density distribution. For illustrative purposess, one can study the dispersion of initally localized wavepacket $P(x,0) = \delta(x)$, and find the behavior of its dispersion
$\sigma(t) = 2\frac{v}{\gamma}\sqrt{\gamma t+e^{-\gamma t}-1}$.
For $\gamma t \ll 1$, the dispersion growths ballistically $\sigma(t) \approx \sqrt{2}vt$. In contrast, at late times, $\gamma t\gg 1$, the wavepacket propagates diffusively,  $\sigma \sim \sqrt{D t}$, where $D= 2v^2/\gamma$ is the diffusion coefficient.

To observe the direct signatures of quasi-hydrodynamic behavior, one needs to find solutions for dimensions higher than one. For higher dimensions, we study an approximate dynamics in which we consider only first-order corrections $\sim v/\gamma$ and neglect the memory effects in Eq.$\,$\eqref{eq:quantum_corrections} (the Born-Markov approximation). First, we consider the equation of motion for the parameters forming $I_k$ in Eq.$\,$\eqref{eq:current_evo},
\be \label{eq:Fs}
\begin{split}
&\frac{\partial F^{\pm}_{kl}}{\partial t}  =  -\gamma F^\pm_{kl}\pm v\nabla_l J_k\pm v\nabla_k J_l + I^{(\pm)}_{kl},\quad k\neq l\\
&\frac{\partial F^+_{kk}}{\partial t}  =  -\gamma F^+_{kk}+v\nabla_k J_k+I^0_{k}
\end{split}
\ee
where $I^{\pm,0}_{kl}$ are terms containing third diagonal elements of the density matrix and satisfying $\Tr(H_0^2I^{\pm,0}_{kl})=0$.
These terms generate correctons of at least second order $\sim v^2/\gamma^2$ to the equation of motion on the current density and can be omitted.  In addition, in the Markov approximation, we neglect the effect of the time derivatives, putting $\partial F^{\pm}_{kl}/\partial t=0 $.
Then, combining Eq.$\,$\eqref{eq:current_evo} and Eq.$\,$\eqref{eq:Fs}, one may derive the approximated equations of motion on the current density
\be\label{eq:approx_dynamics}
\begin{split}
\frac {\partial J_k}{\partial t}  = &-\gamma J_k - 2v^2\nabla_k P  \\
&+ \frac {v^2}{\gamma}\Bigl(\nabla_{k}^2J_k + 2\sum_{k\neq k'}\nabla_{k'}^2J_k+2\nabla_{k}\sum_{k\neq k'}\nabla_{k'}J_{k'}\Bigl)
\end{split}
\ee 
The general expression for the memory function as well as its transition into Eq.$\,$\eqref{eq:approx_dynamics} using the approximation above is derived in \textit{SI: Part 2}.

Combined with the continuity conditions in Eq.$\,$\eqref{eq:continuity}, Eq.$\,$\eqref{eq:approx_dynamics} describes the system in terms of observable local quantities. The solutions converge to exact behavior in the asymptotic limit $v/\gamma\to0$ and applicable for large enough $\gamma$. Surprisingly, in 1D case, these equations coincide with linearized Navier-Stokes equations with effective viscosity $\eta = v^2/\gamma$ and with an added current dissipation term. 
In 2D Eq.$\,$\eqref{eq:approx_dynamics} cannot be reduced to the standard Navier-Stokes equations: although each individual term in Eq. $\,$\eqref{eq:approx_dynamics} takes the same form as the corresponding term in the Navier-Stokes equations, the values of the coefficients of the terms lie outside of the regime for those values permitted by Navier-Stokes. Nevertheless, the last term in Eq.$\,$\eqref{eq:approx_dynamics} is responsible for the emergence of vortices and viscous flow of the particles. Below we demonstrate this for particular physical problems.

\textbf{Special sample geometries}. In the experimental  setting, special sample geometries can reveal and amplify the effect of the dephasing-induced viscosity. To demonstrate this, we start with a standard problem in fluid mechanics and focus on a current flow between two parallel plane surfaces with no-slip boundary conditions \cite{0011-jaggi1991electron}. This problem represents transport in films and wires \cite{0050-molenkamp1994electron} where disorder scattering suppresses the transport at the material surface. 
In cases of ballistic and diffusive regimes, there is no mechanism equilibrating velocity in the bulk and near the surface, therefore the current density should be independent on the position between surfaces. In contrast, a viscous fluid is characterized by Poiseuille parabolic shape of current density originated from the viscous friction between layers moving with different velocities. This setting, therefore, is a good geometry to probe the viscosity of the system.

Below we analyze the problem in the presence of the dephasing using exact numerical simulations for full density operator in Eq.$\,$\eqref{eq:master_equation} (see \textit{SI: Part 5}) and compare it to the analytical approximation for local observables in Eq.$\,$\eqref{eq:continuity} and Eq.$\,$\eqref{eq:approx_dynamics}. Both in numerical simulations and analytical calculations, we look for steady states which satisfy zero current density at the surfaces.
The numerics suggest that  
in the crossover regime, similarly to viscous fluid dynamics, the dephasing effects lead to spatially dependent current density. The plot in Fig. \ref{fig:figure_2}a demonstrates the evolution of the current density shape with the increase of the dephasing rate $\gamma$.
In both ballistic ($\gamma L/v\ll1$) and diffusive ($\gamma L/v\gg1$) limits, the system exhibits space-independent current density. However, in the vicinity of the critical value of the dephasing rate $\gamma_0 = 2\pi v/L$, the profile has distinct maximum in the center between surfaces. This behavior can be explained using the approximate equations of motion in Eq.$\,$\eqref{eq:approx_dynamics} (see \textit{SI: Part 3}). The profile is determined by a competition between the Ohmic dissipation term $\sim \gamma$ and the quasi-viscosity term $\sim v^2/\gamma$ inducing local equilibration of the velocity. The comparison between numerics and the approximation is shown in Fig. \ref{fig:figure_2}b. Remarkably, the local description in Eq.$\,$\eqref{eq:approx_dynamics} gives a quite accurate description of current density for $\gamma\sim\gamma_0$. At the same, in the regimes of small $\gamma$ the flow dynamics are different from the analytic prediction due to the breaking of the approximations used to derive Eq.$\,$\eqref{eq:approx_dynamics}.

To study the problem in more detail, we also look into the flow resistivity of the channel $r$ obtained numerically, which can be compared with analytical expression $r_a$ following from Eq.$\,$\eqref{eq:approx_dynamics} (see \textit{SM: Part 3}),
\be\label{eq:resistivity}
r = - \frac{\partial P}{\partial y} \frac{L}{I_0},\quad r_a = \frac{r_D}{1-\tanh(X)/X},
\ee
where $I_0$ is the total current between surfaces, $r_D = \gamma/2v^2$ is the Drude resistivity, and $X = \gamma L/2\sqrt{2}v$. In nearly equilibrium electronic systems with Fermi energy $\ve_F$, the electrical resistivity $r_{\rm el}$ can be connected to the flow resistivity in Eq.$\,$\eqref{eq:approx_dynamics} as $r_{\rm el} = r/e^2D(\ve_F)$, where $D(\ve)$ is the density of states near Fermi surface. The curves in Fig. \ref{fig:figure_2}c show the ratio of the resistance to Drude result valid for diffusive transport.
Both numerical simulation and analytic approximation predict rapid increase of resistivity in the critical regime. For small $\gamma$, however, the quasi-hydrodynamic result diverges, while the exact numerical result for resistance $r$ remains finite.

Another configuration uncovering the unusual properties of the quantum dissipative regime is a 2D system consisting of the gas of free carriers in a finite size sample (with no-stress boundary conditions) and two short contacts, the source and the drain, attached. This geometry was proposed in \cite{0075-levitov2016electron} as a method to observe the effects of non-local response and vorticity. Here we demonstrate that a similar effect also takes place in systems with dephasing in the critical regime. The panels in Fig. \ref{fig:figure_2}d show numerical simulation using Eq.$\,$\eqref{eq:master_equation} for steady-state density and current density for large and moderate dephasing rate $\gamma$. In the critical regime $\gamma L/v\sim 1$, current backflow appears in the systems leading to vortex formation. As seen in Fig. \ref{fig:figure_2}e, this effect results in sign-changing deviation of probability density near the current leads. In electronic systems, this effect would lead to sign-changing of the voltage across the sample edge $V\sim \Delta P/eD(\ve_F)$ providing detectable experimental evidence of the dephasing-assisted viscous regime. At the same time the analytical calculation using approximated equation of motion in Eq.$\,$\eqref{eq:approx_dynamics} also predicts existence of vortices and the non-local response (see \textit{SI: Part 4}).

\textbf{Discussion}.
In summary, we have demonstrated the existence of a new type of propagation regime in the crossover between quantum coherent transport and diffusion. Environmentally induced quantum viscosity (EQV) is the hydrodynamic behavior of a single
quantum particle or of noninteracting quantum particles, brought on by the 
interplay between wavelike propagation and environmentally induced decoherence.
The fluid-like dynamics of excitations under dissipation can find application in electronics using viscous effects in the absence of strong interactions 
between charge carriers.  EQV provides a potential alternative explanation 
to the results of existing hydrodynamic experiments \cite{0159-crossno2016observation,0153-bandurin2016negative,0020-kumar2017superballistic,0011-molenkamp1994observation}. In addition to solid-state systems such as encapsulated graphene, environmentally assisted quantum fluid dynamics might be observable in cold 
atoms in optical lattices, photonic crystals \cite{joannopoulos2011photonic} 
and photonic simulators \cite{aspuru2012photonic}. Measuring the value of
environmentally induced quantum viscosity should allow the determination
of the strength of interaction between quantum particles and their 
dephasing environment. 

\textit{Acknowledgements:}
Oles Shtanko was supported by an ExxonMobil-MIT Energy Fellowship.  Seth Lloyd
was supported by NSF, AFOSR, and ARO under the Blue Sky Initiative.
\bibliography{bibliography}

\clearpage

\pagebreak

\setcounter{page}{1}
\setcounter{equation}{0}
\setcounter{figure}{0}
\renewcommand{\theequation}{S.\arabic{equation}}
\renewcommand{\thefigure}{S\arabic{figure}}
\renewcommand*{\thepage}{S\arabic{page}}

\onecolumngrid


{\Large
\textbf{Supplementary Information:}}\\


%
%

\textbf{Part 1: Equations of motion}\\

In this section we derive the equations of motion on the probability density and current density using the Eq.$\,$(3) in the main text. 
The dynamics of probability density is connected to the current density as
\be
\begin{split}\label{eq:discrete_cont}
&\frac{\partial P}{\partial t} = -\frac{iv}{a}\Bigl(\rho(\vec r+\vec e_k,\vec r)-\rho(\vec r,\vec r+\vec e_k)\Bigl)-\frac{iv}{a}\Bigl(\rho(\vec r-\vec e_k,\vec r)-\rho(\vec r,\vec r-\vec e_k)\Bigl) = \\
& = -\frac{iv}{a}\Bigl(\rho(\vec r+\vec e_k,\vec r)-\rho(\vec r,\vec r+\vec e_k)\Bigl)-\frac{iv}{a}\Bigl(\rho(\vec r-\vec e_k,\vec r-\vec e_k+\vec e_k)-\rho(\vec r-\vec e_k+\vec e_k,\vec r-\vec e_k)\Bigl) = 
-\sum_k D^{L}_kJ_k
\end{split}
\ee
where the discrete (left) derivative on a function of  vector on the lattice $f(\vec r)$ is defined as $D_k^{L}f(\vec r) = \bigl(f(\vec r)-f(\vec r-\vec e_k)\bigl)/a$. In the continuous limit, $a\to 0$, the discrete derivative transforms into the gradient, $D^{L}_k\to \nabla_k$. In this limit, Eq.$\,$\eqref{eq:discrete_cont} is exactly the continuity condition Eq.(4) in the main text.

The time derivative of the current density vector, in turn, is
\be
\begin{split}
&\frac{\partial J_k}{\partial t} = -\gamma J_k+\frac{v^2}a\sum_l\Bigl(\rho(\vec r+\vec e_k +\vec e_l,r)-\rho(\vec r+\vec e_k,\vec r+\vec e_l)\Bigl)-\frac{v^2}a\sum_l\Bigl(\rho(\vec r+\vec e_l,\vec r+\vec e_k)-\rho(\vec r,\vec r+\vec e_k +\vec e_l)\Bigl)\\
&+\frac{v^2}a\sum_l\Bigl(\rho(\vec r+\vec e_k -\vec e_l,r)-\rho(\vec r+\vec e_k,\vec r-\vec e_l)\Bigl)-\frac{v^2}a\sum_l\Bigl(\rho(\vec r-\vec e_l,\vec r+\vec e_k)-\rho(r,\vec r+\vec e_k -\vec e_l)\Bigl)
\end{split}
\ee
We rearrange this expression separating the terms containing diagonal and off-diagonal entries in the density matrix only, and then combine them into groups such as
\be
\begin{split}
&\frac{\partial J_k}{\partial t}=-\gamma J_k-\frac{2v^2}a\Bigl(\rho(\vec r+\vec e_k,\vec r+\vec e_k)-\rho(\vec r,\vec r)\Bigl)\\
&+\frac{v^2}a\sum_l\Bigl(\rho(\vec r+\vec e_k +\vec e_l,\vec r)+\rho(\vec r,\vec r+\vec e_k +\vec e_l)\Bigl)-\frac{v^2}a\sum_l\Bigl(\rho(\vec r-\vec e_l+\vec e_k +\vec e_l,\vec r-\vec e_l)+\rho(\vec r-\vec e_l,\vec r-\vec e_l+\vec e_k +\vec e_l)\Bigl)\\
&+\frac{v^2}a\sum_{l\neq k}\Bigl(\rho(\vec r+\vec e_k -\vec e_l,\vec r)+\rho(\vec r,\vec r+\vec e_k -\vec e_l)\Bigl)-\frac{v^2}a\sum_{l\neq k}\Bigl(\rho(\vec r+\vec e_l+\vec e_k -\vec e_l,\vec r+\vec e_l)+\rho(\vec r+\vec e_l,\vec r+\vec e_l+\vec e_k -\vec e_l)\Bigl)
\end{split}
\ee
One can rewrite this expression in the compact form
\be
\frac{\partial J_k}{\partial t}= -\gamma J_k-2v^2 D^R_kP+vD^L_kF_{k}+v\sum_{l\neq k} D^L_lF^+_{kl}-v\sum_{l\neq k} D^R_lF^-_{kl}
\ee
where  $D_k^{L}f(\vec r) = \bigl(f(\vec r)-f(\vec r-\vec e_k)\bigl)/a$ is the discrete (right) derivative, and we used the short notations for the first order correlation functions $
F_{kl}^\pm = v\Bigl(\rho(\vec r+\vec e_k \pm\vec e_l,\vec r)+\rho(\vec r,\vec r+\vec e_k \pm \vec e_l)\Bigl)$, $F_k = F^{+}_{kk}$. In the continuous limit, this expression transforms into Eq.(5) in the main text.

Similar algebra can be used to express the dynamics of parameters $F^\pm_{kl}$. As follows from Eq.$\,$(3) in the main text,
\be
\begin{split}
\frac{\partial F^\pm_{kl}}{\partial t} = -\gamma F^\pm_{kl}-\frac{iv^2}a\sum_m\Bigl(\rho(\vec r+\vec e_k \pm\vec e_l + \vec e_m,\vec r)-\rho(\vec r+\vec e_k\pm\vec e_l,\vec r+\vec e_m, )\Bigl)\\
-\frac{iv^2}a\sum_m\Bigl(\rho(\vec r + \vec e_m,\vec r+\vec e_k \pm\vec e_l)-\rho(\vec r,\vec r+\vec e_k\pm\vec e_l +\vec e_m)\Bigl)\\
-\frac{iv^2}a\sum_m\Bigl(\rho(\vec r+\vec e_k \pm\vec e_l - \vec e_m,\vec r)-\rho(\vec r+\vec e_k\pm\vec e_l,\vec r-\vec e_m )\Bigl)\\
-\frac{iv^2}a\sum_m\Bigl(\rho(\vec r - \vec e_m,\vec r+\vec e_k \pm\vec e_l)-\rho(\vec r,\vec r+\vec e_k\pm\vec e_l -\vec e_m)\Bigl)
\end{split}
\ee
Let us consider first the expression for $F^+_{kl}$ in the case $k\neq l$. Similar to the equations for current density, we separate terms containing different orders of correlation in the density matrix
\be
\begin{split}
\frac{\partial F^+_{kl}}{\partial t} = -\gamma F^+_{kl}&+\frac{iv^2}a\Bigl(\rho(\vec r+\vec e_l+\vec e_k,\vec r+\vec e_l)-\rho(\vec r+\vec e_l,\vec r+\vec e_l+\vec e_k)\Bigl)-\frac{iv^2}a\Bigl(\rho(\vec r+\vec e_k,\vec r)-\rho(\vec r,\vec r+\vec e_k)\Bigl)\\
&+\frac{iv^2}a\Bigl(\rho(\vec r+\vec e_k+\vec e_l,\vec r+\vec e_k)-\rho(\vec r+\vec e_k,\vec r+\vec e_k+\vec e_l)\Bigl)-\frac{iv^2}a\Bigl(\rho(\vec r+\vec e_l,\vec r)-\rho(\vec r,\vec r+\vec e_l)\Bigl)\\
&-\frac{iv^2}a\sum_{m}\Bigl(\rho(\vec r+\vec e_k +\vec e_l + \vec e_m,\vec r)-\rho(\vec r,\vec r+\vec e_k +\vec e_l + \vec e_m)\Bigl)\\
&+\frac{iv^2}a\sum_{m}\Bigl(\rho(\vec r- \vec e_m+\vec e_k +\vec e_l + \vec e_m,\vec r- \vec e_m)-\rho(\vec r- \vec e_m,\vec r- \vec e_m+\vec e_k +\vec e_l + \vec e_m)\Bigl)\\
&-\frac{iv^2}a\sum_{m\neq k,l}\Bigl(\rho(\vec r+\vec e_k +\vec e_l - \vec e_m,\vec r)-\rho(\vec r,\vec r+\vec e_k +\vec e_l - \vec e_m)\Bigl)\\
&+\frac{iv^2}a\sum_{m\neq kl}\Bigl(\rho(\vec r+ \vec e_m+\vec e_k +\vec e_l - \vec e_m,\vec r+ \vec e_m)-\rho(\vec r+ \vec e_m,\vec r+ \vec e_m+\vec e_k +\vec e_l - \vec e_m)\Bigl)
\end{split}
\ee
This expression can be combine into the compact form
\be\label{eq:dFdt_plus}
\frac{\partial }{\partial t} F^+_{kl}(\vec r,t)= -\gamma F^+_{kl}(\vec r,t)+vD_l^RJ_k(\vec r,t)+vD_k^RJ_l(\vec r,t)-\sum_{m}vD_m^LF^{++}_{klm}(\vec r,t)+\sum_{m\neq k,l}vD_m^RF^{+-}_{klm}(\vec r,t), \qquad k\neq l 
\ee
where $F_{klm}^{\pm\pm} = iv\Bigl(\rho(\vec r+\vec e_k \pm\vec e_l\pm\vec e_m,\vec r)-\rho(\vec r,\vec r+\vec e_k \pm \vec e_l\pm\vec e_m)\Bigl)$. 

In the similar fashion, one may obtain the dynamic equations for the case $k=l$,
\be
\frac{\partial F_{k}}{\partial t} = -\gamma F_{k}+vD_k^RJ_k-\sum_{m}vD_m^LF^{++}_{kkm}+\sum_{m\neq k}vD_m^RF^{+-}_{kkm}
\ee
as well as for the
\be\label{eq:dFdt_minus}
\frac{\partial }{\partial t} F^-_{kl}(\vec r,t)= -\gamma F^-_{kl}(\vec r,t)-vD_l^LJ_k(\vec r,t)-vD_k^RJ_l(\vec r-\vec e_l)+\sum_{m\neq k}vD_m^RF^{--}_{klm}(\vec r,t)-\sum_{m\neq l}vD_m^LF^{-+}_{klm}(\vec r,t), \qquad k\neq l 
\ee
The expressions in Eqs.\eqref{eq:dFdt_plus}-\eqref{eq:dFdt_minus} result in Eq.(8) in the main text in the continuous limit.\\

\textbf{Part 2: The memory function}\\

\textbf{General formula}. To derive the memory function, we first rewrite the dynamics of the system in terms of variables representing conserved quantities. For example, in the ballistic regime, i.e. $\gamma=0$, the integral over lattice volume for each of probability density and current density are a conserved quantity. Additionally, thermodynamic coherent quantum systems obey infinite number of conserved quantities scaling linearly with the system volume. As we show below, it is possible to express the full set of conserved quantities as
\be\label{eq:ioms}
F_{\vec n} = \int_V d\vec r f_{\vec n} (\vec r,t),\qquad f_{\vec n}(\vec r,t) = i^{c(\vec n)} v\Bigl(\rho(\vec r,\vec r+\vec n)+(-1)^{c(\vec n)} \rho(\vec r+\vec n,\vec r)\Bigl),
\ee
where $\vec n$ is $d$-dimensional vector of integers, and
\be
\vec n = \sum n_k\vec e_k, \qquad c(\vec n) = \sum_k n_k
\ee

Here and below we refer to $f_{\vec n}(\vec r,t)$ as quantum correlation function of order $\eta(\vec n) = \sum_k |n_k|$.
It is straightforward to see that the correlation function of zeroth order $\eta=0$ is nothing but the probability density up to the factor, $P(\vec r,t) =f_0(\vec r,t)/2v$. Similarly, the current density components $J_k(\vec r,t)$ are exactly the correlation functions of order $\eta=1$. 
Below we show that dynamics of general correlation functions $f_{\vec n} (\vec r,t)$ is described by a set of continuity equations connecting high orders to lower ones. At the next step, we exclude the high order correlation functions from the equations of motion, leaving the equations on the probability density and the current density only. The latter method works for a system in the presence of the dephasing too; therefore, from the start, we consider the general case $\gamma\neq0$.

 To derive the equations of motion, we use Eq.$\,$(3) in the main text. Taking the time derivative simultaneously of the left- and the right-hand side of the second expression in Eq.$\,$\eqref{eq:ioms}, we substitute the time derivatives of density operators from Eq.$\,$(3) in the main text, and obtain
\be
\begin{split}
\frac{\partial}{\partial t}f_{\vec n}(\vec r,t) = -\gamma f_{\vec n}(\vec r,t) -\frac{iv}a i^{c(\vec n)}\sum_{k=1}^d\sum_{\xi=\pm1}\Bigl(&\rho(\vec r +\xi\vec e_k,\vec r+\vec n)+(-1)^{c(\vec n)} \rho(\vec r+\vec n+\xi\vec e_k,\vec r )\\
&-\rho(\vec r ,\vec r+\vec n+\xi\vec e_k)-(-1)^{c(\vec n)} \rho^\dag(\vec r+\vec n,\vec r+\xi\vec e_k)\Bigl).
\end{split}
\ee
To simplify the analysis, one can use the mixed Fourier-Laplace transformation for space and time respectively, and rewrite the equation for corresponding transformation components,
\be
f_{\vec n}(\vec q,s) = \sum_{\vec r}\int_0^\infty dt f_{\vec n}(\vec r,t)e^{st-i\vec q\vec r},
\ee
To perform the Laplace transform, we need to specify correlation function values at the initial moment of time, $t=0$. Without a loss of generality, we assume that the particle density is $P(\vec r,t=0) = P_0(\vec r)$ and the system has no quantum correlations initially,
\be \label{eq:initial_qc}
 f_{\vec n}(\vec r,t=0) = 0\quad{\rm if}\quad \eta(\vec n)\geq 1,
\ee
where the expression in Eq.$\,$\eqref{eq:initial_qc} also includes the current density.

In the Fourier-Laplace representation,  
after rearranging, the equations receive their compact form,
\be
sf_{\vec n}(\vec q,s)= -\gamma f_{\vec n}(\vec q,s)-v\sum_k \Bigl(D_{q_k}f_{\vec n+\vec e_k}(\vec q,s)-D^*_{q_k}f_{\vec n-\vec e_k}(\vec q,s)\Bigl),\qquad D_{q_k}= \frac 
1{a}(e^{iq_ka}-1),
\ee
In the limit $a\to0$, the Fourier transform of discrete derivative $D_{q_k}$ obtains its continuous form, $D_{q_k}\to iq_k$. Adding the continuity equation for the density (Eq.(4) in the main text), we obtain the set of equation describing local variables and high order correlation functions together,
\be\label{eq:equations_of_motion}
\begin{cases}
sP(\vec q,s) = P_0(\vec q)-i\sum_m q_mJ_m(\vec q,s)\\
sJ_k(\vec q,s) = -\gamma J_k(\vec q,s)-2iv^2 q_k P(\vec q,s)-iv\sum_{k'} q_{k'}f_{\vec e_k+\vec e_{k'}}(\vec q,s)-iv\sum_{k'\neq k} q_{k'}f_{\vec e_k-\vec e_{k'}}(\vec q,s),\\
sf_{\vec n}(\vec q,s) = -\gamma f_{\vec n}(\vec q,s)-iv\sum_{m} q_m\Bigl( f_{\vec n+\vec e_m}(\vec q,s)+ f_{\vec n-\vec e_{m}}(\vec q,s)\Bigl),\quad \eta(\vec n)\geq2
\end{cases}
\ee
where $P_0(\vec q)$ is the Fourier transform of the initial at the time $t=0$. This set of equation gives an equivalent description of the system compared with the density operator.

 To integrate out the high order correlations in Eq.$\,$\eqref{eq:equations_of_motion}, we need to express $f_{\vec e_k+\vec e_k'}(\vec q,s)$ and $f_{\vec e_k-\vec e_k'}(\vec q,s)$ as a function of the current density.  The expression in a compact form can be obtained if we
 associate with each variable $f_{\vec n}(\vec q,s)$ a position on a square lattice $|\vec n\>$, and rewrite the last expression in Eq.$\,$\eqref{eq:equations_of_motion} as
\be\label{eq:operator_form}
(s+\gamma+iv\vec q\cdot\vec T)|f(\vec q,s)\> =-iv\sum_{k,m}q_mJ_k(\vec q,s)|\phi_{km}\>
\ee
where $|f(\vec q,s)\> = \sum_{\vec n}f_{\vec n}(\vec q,s)|\vec n\>$, the state $
|\phi_{km}\> = |\vec e_k+\vec e_m\>+|-(\vec e_k+\vec e_m)\>+\Bigl(|\vec e_k-\vec e_m\>+|\vec e_m-\vec e_k\>\Bigl)(1-\delta_{km})$, the short notation is $\vec q\cdot\vec T = \sum_m q_m T_m$,
and matrices $T_m$ are defined as
\be
 T_m = \sum_{\vec n,\vec n'\in S} |\vec n'\>\<\vec n+\vec e_m|+{\rm h.c.},\qquad {\rm where}\qquad \eta(\vec n),\eta(\vec n')>1
\ee
Using Eq.$\,$\eqref{eq:operator_form}, we can express the variables $f_{\vec e_k\pm\vec e_{m}}$ as
\be
f_{\vec e_k\pm\vec e_{m}}(\vec q,s) =\<\vec e_k\pm\vec e_m|f(\vec q,s)\> = -iv\sum_{m',k'=1}^d\<\vec e_k\pm\vec e_m|\frac1{s+\gamma+iv\vec q\cdot\vec T}|\phi_{k'm'}\>q_{m'}J_{k'}(\vec q,s) 
\ee

This expression, after substitution in Eq.$\,$\eqref{eq:equations_of_motion}, results in the equation on the current density,
\be\label{eq:approx_dyn_fourier}
sJ_k(\vec q,s) = -\gamma J_k(\vec q,s)-2iv^2 q_k P(\vec q,s)+\sum_{k'}K_{kk'}(\vec q,s)J_{k'}(\vec q,s),
\ee
where the memory function is
\be\label{eq:memory_function}
K_{kk'}(\vec q,s) = -v^2\sum_{mm'}\<\psi_{km}|\frac{q_mq_{m'}}{s+\gamma+iv\vec q\cdot\vec T}|\phi_{k'm'}\>
\ee
where $|\psi_{km}\> = |\vec e_k+\vec e_{m}\>+|\vec e_k-\vec e_{m}\>(1-\delta_{km})$.

Let us demonstrate now that this general result reduces to Eq.$\,$(9). To obtain the Markov approximation, we neglect the dependence on $s$. Then, let us expand the expression in Eq.$\,$\eqref{eq:memory_function} and take only first order in $v/\gamma$ (Born approximation). As result, we obtain
\be
K_{kk'}(\vec q,s) = -\frac{v^2}{\gamma}\sum_{mm'}q_mq_{m'}\<\psi_{km}|\phi_{k'm'}\> + O\Bigl(\frac1{\gamma^2}\Bigl)= -\frac{v^2}{\gamma}\Bigl(\delta_{kk'}\Bigl(q_k^2+2\sum_{m\neq k}q_{m}^2\Bigl)+2(1-\delta_{kk'})q_kq_{k'}\Bigl)+O\Bigl(\frac1{\gamma^2}\Bigl)
\ee
Inserting this expression in Eq.$\,$\eqref{eq:approx_dyn_fourier}, we obtain the result which coincides with our result in Eq.(9) in the main text.\\

\textbf{Memory function in 1D}. Let us consider the case of 1D system. In this case we can use a single integer index $n$ for indexing correlation functions, and also a single wavector component $q_x\equiv q$. The simplicity of the structure of operator $q_xT_x$ in 1D allows to obtain the analytical expression for the memory function in Eq.$\,$\eqref{eq:equations_of_motion}. To do this explicitly, let us rewrite the set of 1D continuity equations in Eq.$\,$\eqref{eq:equations_of_motion},
\be\label{eq:1d_hierarchy}
sf_n(q,s) = -\gamma f_n(q,s) -iv q\Bigl(f_{n-1}(q,s)+f_{n+1}(q,s)\Bigl),
\ee
into the form of a recurrence equation,
\be\label{eq:recurrence_cond}
f_{n+1}(q,s) = -ivq\mathcal X(q,s) f_{n}(q,s), \qquad n\geq 1.
\ee
where $X(q,s)$ is a function which assumed to not depend on the index n. 
Inserting this condition into Eq.\eqref{eq:1d_hierarchy}, and rearranging it in the form of recurrence condition,
\be\label{eq:explicit_recurrence_cond}
f_n(q,s) = \frac {-ivq}{s+\gamma +v^2q^2\mathcal X(q,s)} f_{n-1}(q,s)
\ee
we get another recurrence equation which needs to be compared with Eq.\eqref{eq:explicit_recurrence_cond}. The consistency require the function $\mathcal X(q,s)$ to satisfy a quadratic equation,
\be
v^2q^2\mathcal X^2+(s+\gamma)\mathcal X-1=0
\ee
This expression has two possible solutions,
\be\label{eq:xi_solutions}
\mathcal X(q,s) = \frac 1{v^2q^2}\Biggl(- \frac{s+\gamma}2 \pm \sqrt{\frac{(s+\gamma)^2}4+v^2q^2}\Biggl).
\ee
The physical solution must satisfy $X(q,s)\to0$ as $q\to0$, since transport is vanishing in absence of gradients. Therefore we need to choose plus sign in Eq.$\,$\eqref{eq:xi_solutions}. Using expression for $X(q,s)$, one can express the memory function as
\be
K(q,s) = - \frac{s+\gamma}2 + \sqrt{ \frac{(s+\gamma)^2}4+v^2 q^2}
\ee
To illustrate the transition between coherent and diffusive transport, we use the memory function to derive the Green's function $G(x,t)$ for density defined as
\be
P(x,t) = \int_{-\infty}^\infty dx' G(x-x',t)P(x',0)
\ee
In the Fourier-Laplace space,
\be\label{eq:1d_greens_function}
G(q,s) =\frac 1{s+2K(q,s)} = \frac1{\sqrt{(s+\gamma)^2+4v^2q^2}-\gamma}
\ee
In the limit of strong dephasing, $\gamma\to\infty$, this expression reduces to Green's function for diffusive motion, while for vanishing dephasing, $\gamma\to0$ it has a lightcone with velocity $v$,
\be
\lim_{\gamma\to0}G(q,s) =\frac1{\sqrt{s^2+4v^2q^2}},\qquad \lim_{\gamma\to \infty}G(q,s) = \frac{1}{s+2\frac{v^2}\gamma q^2}
\ee 
The plot of the Green's function is shown in Fig. 1b in the main text.

\begin{figure*}[t]
\includegraphics[width=0.7\columnwidth]{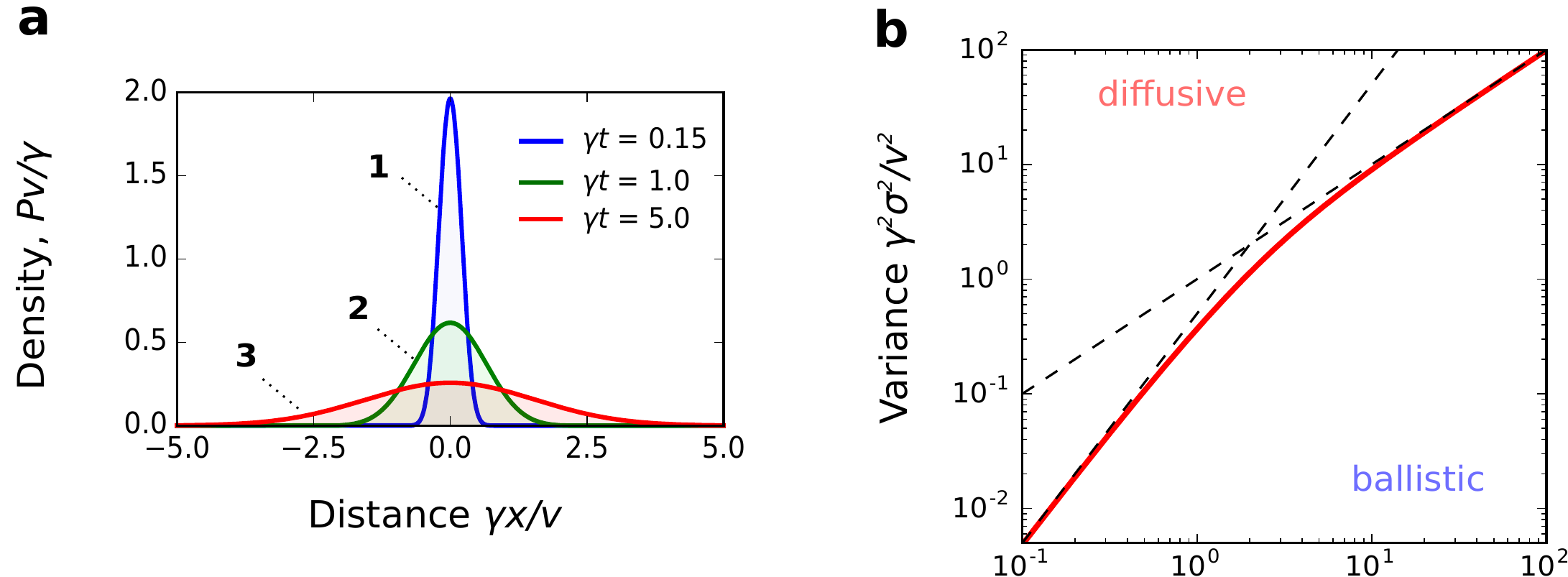}
\caption{\textbf{Wavepacket propagation in the presence of the dephasing}. \textbf{a}, Time dependence of the density deistrubution for a wavepacket for $v=1$, $\gamma=1$ with the initial box-shaped profile $P(x,0)=0.5$, $x\in[-1,1]$. \textbf{b}, The dispersion $\sigma^2(t)$ as function of time for initially localized wavepacket $P(x,0)=\delta(x)$. In logarithmic scale, the crossover is evident between the crossover from ballistic regime $\sigma^2\sim t^2$ to diffusive regime $\sigma^2\sim t$. }
\label{figs:wavepacket_spreading}
\end{figure*}

In terms of propagation, it is instructive to consider a wavepacket localized at initial time, $P(x,0) = \delta(x)$. The width of the wavepacket $\sigma(t)$ can be defined as
\be
\sigma^2(t) = \int_{-\infty}^\infty x^2 G(x,t)dx = -\int_{c-i\infty}^{c+i\infty}ds e^{-st}\frac{\partial^2}{\partial q^2}G(q,s)\biggl|_{q=0}
\ee
Inserting Eq.$\,$\eqref{eq:1d_greens_function} into this expression, we obtain
\be
\sigma^2(t) = \frac{4v^2}{\gamma^2}\Bigl(\gamma t+e^{-\gamma t}-1\Bigl)
\ee
As discussed in the main text, this expression shows ballistic behavior for short times and diffusive dynamics at the later times,
\be
\lim_{t\to0}\sigma(t)\sim vt,\qquad \lim_{t\to\infty}\sigma(t)\sim\sqrt{Dt}
\ee
where $D = 2\gamma^2/v$ is the diffusion coefficient and $u = v$ is wavepacket ballistic spreading velocity. At the same time the velocity of the wavepacket ``lightcone'' is $u'=2v$ (see Fig. 1b in the main text). The behavior of the wavepacket dispersion is illustrated in Fig. \ref{figs:wavepacket_spreading}.\\

\textbf{Part 3: Flow Between Two Parallel Surfaces}\\

\textbf{General solution}. In the main text, we consider a problem of a steady flow between two parallel infinite flat surfaces in a presence of the dephasing. We assume that the surfaces obeys the no-slip boundary conditions originated from the microscopic roughness of the surfaces. These boundary conditions require that the set of equations in Eq.(4) and Eq.(9) in the main text, with the steady state condition $\frac{\partial}{\partial t} P=0$ and $\frac{\partial}{\partial t} \vec J=0$, needs to be combined with the condition $J_y(\pm L/2,y)=0$, where we set the surface positions at $x=\pm L/2$. To solve the problem, we exploit the conventional method in electrostatics and consider the problem in presence of sources $\vec S = (S_0,S_1,S_2)$. The sources must be added to the system such that equations
\be
\begin{cases}
- \nabla \cdot \vec J = S_0,\\
-\gamma J_k - 2v^2\nabla_k P 
+ \frac {v^2}{\gamma}\Bigl(\nabla_{k}^2J_k + 2\sum_{k\neq k'}\nabla_{k'}^2J_k+2\nabla_{k}\sum_{k\neq k'}\nabla_{k'}J_{k'}\Bigl)
= S_k
\end{cases}
\ee
satisfy the boundary conditions automatically. For the problem we solve in this section, the sources need be chosen as
\be\label{eq:sources}
\vec S = \biggl(A\lim_{y_\pm\to\pm\infty}\Bigl(\delta(y-y_-)-\delta(y-y_+)\Bigl),0,B\sum_{k=-\infty}^\infty\delta\left(x+L/2+kL\right)\biggl)
\ee
where $A$ and $B$ are constants we will determine later. The first component of the vector $\vec S$ represent distant a source and a drain in the system generating the flow, while the the third component corresponds to the dissipation of current at surfaces resulting in the no-slip boundary conditions. In our problem, we focus only on the part of the space enclosed in $-L/2<x<L/2$.

It is convenient to use the linearity of the equation of motion, and decompose the result into two independent solutions corresponding to each of vector components in Eq.\eqref{eq:sources},
\be\label{eq:general_sol}
P(\vec r) = P_1(\vec r)+P_2(\vec r),\qquad \vec J(\vec r) = \vec J_1(\vec r)+\vec J_2(\vec r).
\ee
The first solution, $P_1$ and $\vec J_1$, we take as a homogeneous solution for the problem with the source term $\vec S_1 = (A\lim_{y_\pm\to\pm\infty}\Bigl(\delta(y-y_-)-\delta(y-y_+)\Bigl),0,0)$. One may check that this solution has the form
\be
P_1(\vec r) = -\gamma Ay/2v^2,\qquad J_{1x}(\vec r) = 0,\qquad J_{1y}(\vec r) =A.
\ee
The second solution correspond to the source term $\vec S_2 = (0,0, B\sum_{k=-\infty}^\infty\delta\left(x+L/2+kL\right))$. To express the solution, one may use the Fourier space representation of the equations as
\be
\mathcal L \vec \Psi_2(\vec q) = \vec S_2(\vec q),\qquad
\mathcal L = -\left( \begin{matrix}
 0 & iq_x & 0\\
2iv^2q_x &\gamma &0 \\
0 & 0& \gamma(1 +2\lambda^2q_x^2)
\end{matrix}\right),
\qquad \vec S_2(\vec q) = \left(0,0,\frac{2B}{L}\cos(q_xL/2)\right),
\ee
where we have already taken into consideration
 that $\nabla \cdot \vec J=0$ everywhere and problem is homogeneous in $y$-direction (which is equivalent to putting $q_y=0$). Here $\vec \Psi_2  \equiv (P_2(\vec q),J_{2x}(\vec q),J_{2y}(\vec q))$ is the vector of Fourier components, $\lambda = v/\gamma$ is the parameter of length determining the mean free path in the system, and $q_x = 2\pi n/L$, $n\in \mathbb{Z}$ is wavevector taking discrete values. The solution can obtained as 
$\vec \Psi_2 =  \mathcal L^{-1}\vec S_2$,
where the inverse operator is
\be \label{eq:inverse_L}
\mathcal L^{-1} =-\left(
\begin{array}{ccc}
 \gamma q^{-2}_x & -iq^{-1}_x & 0 \\
 -iq^{-1}_x & 0 & 0 \\
 0 & 0 & \gamma^{-1}\bigl(1 + 2\lambda^2q_x^2\bigl)^{-1}  \\
\end{array}
\right).
\ee
Using the matrix elements from Eq.$\,$\eqref{eq:inverse_L}, one can derive the second solution as
\be\label{eq:solution2}
P_2(\vec r) = 0,\qquad J_{2x}(\vec r)=0,\qquad 
J_{2y}(\vec r) =\frac{2B}{\gamma L}\sum_{q_x} \frac{\cos({q_x L/2})}{1+2\lambda^2q_x^2}\exp(iq_x x)
\ee
The boundary conditions are satisfied if
\be
A + \frac{B}{\sqrt{2} \gamma \lambda} {\rm coth}\left(\frac{L}{2\sqrt{2} \lambda}\right) = 0.
\ee
Combined together, the general solution in Eq.$\,$\eqref{eq:general_sol} has the form
\be
P(\vec r) = -\frac{\gamma I_0y}{2v^2\biggl(L-2\sqrt{2} \lambda \tanh\left(\frac{L}{2\sqrt{2} \lambda}\right)\biggl)},\qquad J_{x}(\vec r) = 0,\qquad 
J_{y}(\vec r) =I_0\frac{\biggl(L-2\sqrt{2} \lambda \tanh\left(\frac{L}{2\sqrt{2} \lambda}\right)\sum_{q_x} \frac{\cos({q_x L/2})}{1+\lambda^2q_x^2}\exp(iq_x x)\biggl)}{\biggl(L-2\sqrt{2} \lambda \tanh\left(\frac{L}{2\sqrt{2} \lambda}\right)\biggl)}
\ee
where $I_0$ is the total current through the sample.

The normalized current density is shown in Fig. 2b by solid and dashed curves compared to numerical simulations (shown in dots). Notably, the Born approximation used to derive Eq.(9) predicts the current density profile pretty well even for not too high values of $\gamma L/v\sim 1$.\\

\textbf{Diffusive and projected ballistic limits}. The current density has no simple analytic form for generic values of $\lambda$. However, one can calculate its behavior in asymptotic limits $\lambda\to\infty$ and $\lambda\to0$. The first limit correspond to diffusive transport regime and must converge to exact solution of Eq.$\,$(3) in the main text with the increase of $\lambda$. The limit $\lambda\to0$ describes a projected behavior of the system for a weak dephasing. This solution can be sufficiently different from the real system behaviour since it violates the assumptions used to derive Eq.$\,$(9). Nevertheless, it is interesting to calculate this limit to compare it with the hydrodynamic solution.

In the diffusive limit, $2\pi\lambda/ L \ll 1$, we may replace the sum by integral in Eq.\eqref{eq:solution2},
\be
\sum_{s=\pm1,q_x} \frac{e^{iq_x (x+sL/2)}}{1+2\lambda^2q_x^2} \approx 
 \sum_{s=\pm1}\int_{-\infty}^{\infty} \frac{dq_x}{2\pi} \frac{e^{iq_x (x+sL/2)}}{1+2\lambda^2q_x^2}
=\frac{1}{2\sqrt{2}\lambda}\Bigl( e^{(x-L/2)/\sqrt{2}\lambda}+e^{-(x+L/2)/\sqrt{2}\lambda}\Bigl)
\ee
resulting in the expression for the current density
\be
J_y(\vec r)= I_0\frac{\Bigl(1-2e^{-L/2\sqrt{2} \lambda}\cosh (x/\sqrt{2} \lambda)\Bigl)}{L+2\sqrt{2}\left(e^{-L/\sqrt{2} \lambda }-1\right)  \lambda/L }
\ee
The effects of the non-slip boundary are exponentially suppressed in the bulk, propositional to the probability of collisionless propagation for the equivalent distance to the surface.

In the limit, $2\pi\lambda/L\gg 1$, the denominator is approximately equal to $2\lambda^2q_x^2$ for all $q_x\neq 0$
\be
\sum_{s=\pm1,q_x} \frac{e^{iq_x (x+sL/2)}}{1+2\lambda^2q_x^2} \approx 1 + \frac{ L^2}{8\pi^2 \lambda^2}\biggl(\sum_{n=1}^\infty\frac {(-1)^n}{n^2}e^{2\pi i  n x/L}+{\rm h.c}\biggl)
=1 + \frac{L^2}{4\lambda^2}B_2(\frac xL+\frac 12)
\ee
where $B_2(x) = x^2-x+\frac 16$ is Bernoulli polynomial. Inserting this expression into Eq.$\,$\eqref{eq:solution2}, we derive
\be
J_{y}(\vec r) = \frac{6I_0}L\Bigl(\frac{1}4-\frac{x^2}{L^2}\Bigl)
\ee
This expression coincides with Poiseuille solution in fluid dynamics.\\

\textit{Resistivity}. -- Let us define the specific resistivity of the channel as
\be
r = -\frac{\partial P}{\partial y} \frac L{I_0} =\frac{\gamma/2v^2}{1-\tanh(X)/X}, \qquad X = L/2\sqrt{2}\lambda
\ee
In the corresponding asymptotic limits,
\be
r = \begin{cases}
r_D,\qquad &\lambda/L\ll1, \quad \text{Drude law}\\
12r_D\lambda^2/L^2,\qquad &\lambda/L\gg1, \quad \text{Hagen-Poiseuille flow}
\end{cases}
\ee
where $r_D=\gamma/2v^2 = 1/D$ is the Drude resistivity, $D$ is the diffusion coefficient (see Part 2).
\\

\textbf{Part 4: Flow Between Two Contacts}\\

\begin{figure*}[t]
\includegraphics[width=0.95\columnwidth]{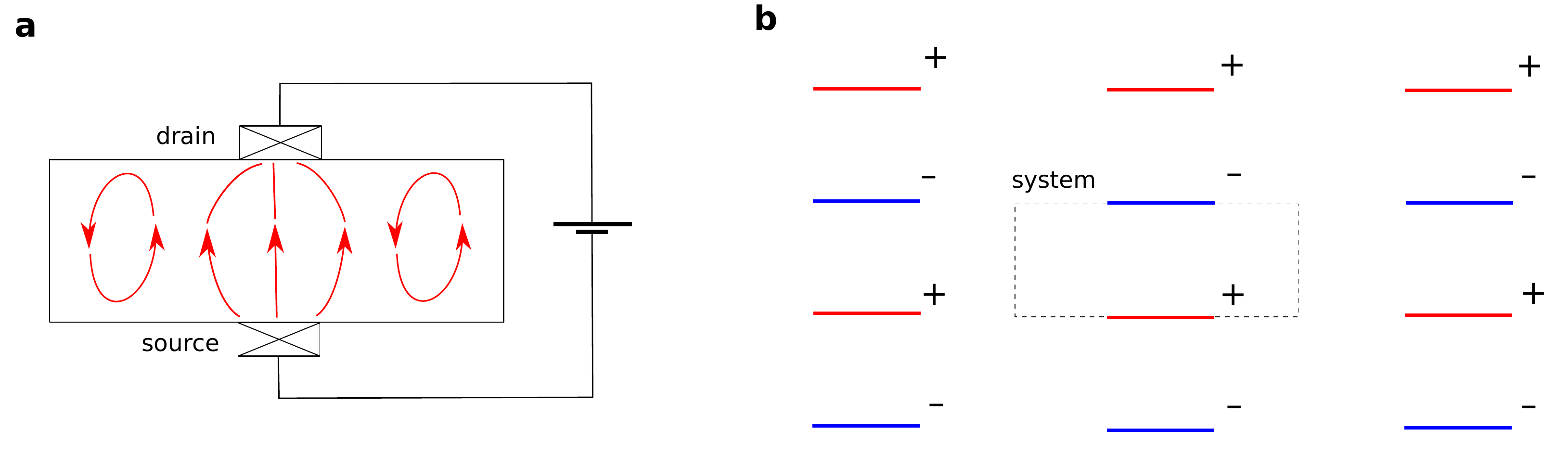}
\caption{\textbf{Flow between two contacts}. \textbf{a}, Schematics of the system. External source drive the flow of carriers through the system via two contacts. \textbf{b}, Source configuration satisfying the boundary conditions. Red regions represent source of particles, blue regions is the drain. Dashed line shows the boundaries of the system.}
\label{figs:two_contacts}
\end{figure*}

In the main text, we study a sample settings of quantum particles propagating in a 2D sample with space dimensions $L_x\times L_y$.  The current flows between two contacts being a source and a drain (see Fig.$\,$\ref{figs:two_contacts}). This configuration implies the boundary conditions on the current density
\be
J_x(\pm L_x/2,y) = 0,\quad 
J_y(x,sL_y/2) = 
\begin{cases}sI_0, \quad &|x|<w/2\\
0, \quad &|x|\geq w/2
\end{cases}
\ee
where $s=\pm 1$ and $w$ is the width of the current leads. The first condition is equivalent to no-stress boundary parallel to $y$ axis. The second condition describes the no-stress boundary conditions parallel to $x$ axis as well as current flow from the source lead to the drain lead. Using these conditions, we assume the homogeneous distribution of current density $J_y = \pm I_0/w$ along the leads which is convenient for calculations.

Applying the method of sources we developed in the previous section, we solve this problem by adding the source term $\vec S$. In particular, we consider the source only for particle density
\be
\vec S = (\Phi(x,y),0,0)
\ee
where the function $\Phi(x,y)$ is periodic in space, $\Phi(x,y+2L_y)=\Phi(x,y)$ and $\Phi(x+L_x,y)=\Phi(x,y)$, and within one period $-L_x/2<x<L_x/2$ it satisfies
\be
\Phi(x,y)=\begin{cases}
\frac{1}{2w}I_0\Bigl(\delta(y-L_y/2)-\delta(y+L_y/2)\Bigl) ,\qquad &|x|\leq w/2\\
0,\qquad &|x|>w/2
\end{cases}
\ee
where the factor $1/2$ appears in the first line due to the geometry of chosen boundary conditions; it contains two mirror-reflected boxes as shown in Fig. $\,$\ref{figs:two_contacts}b. We focus in the space region $-L_x/2<x<L_x/2$ and $-L_y/2<y<L_y/2$.

\begin{figure*}[t]
\includegraphics[width=0.55\columnwidth]{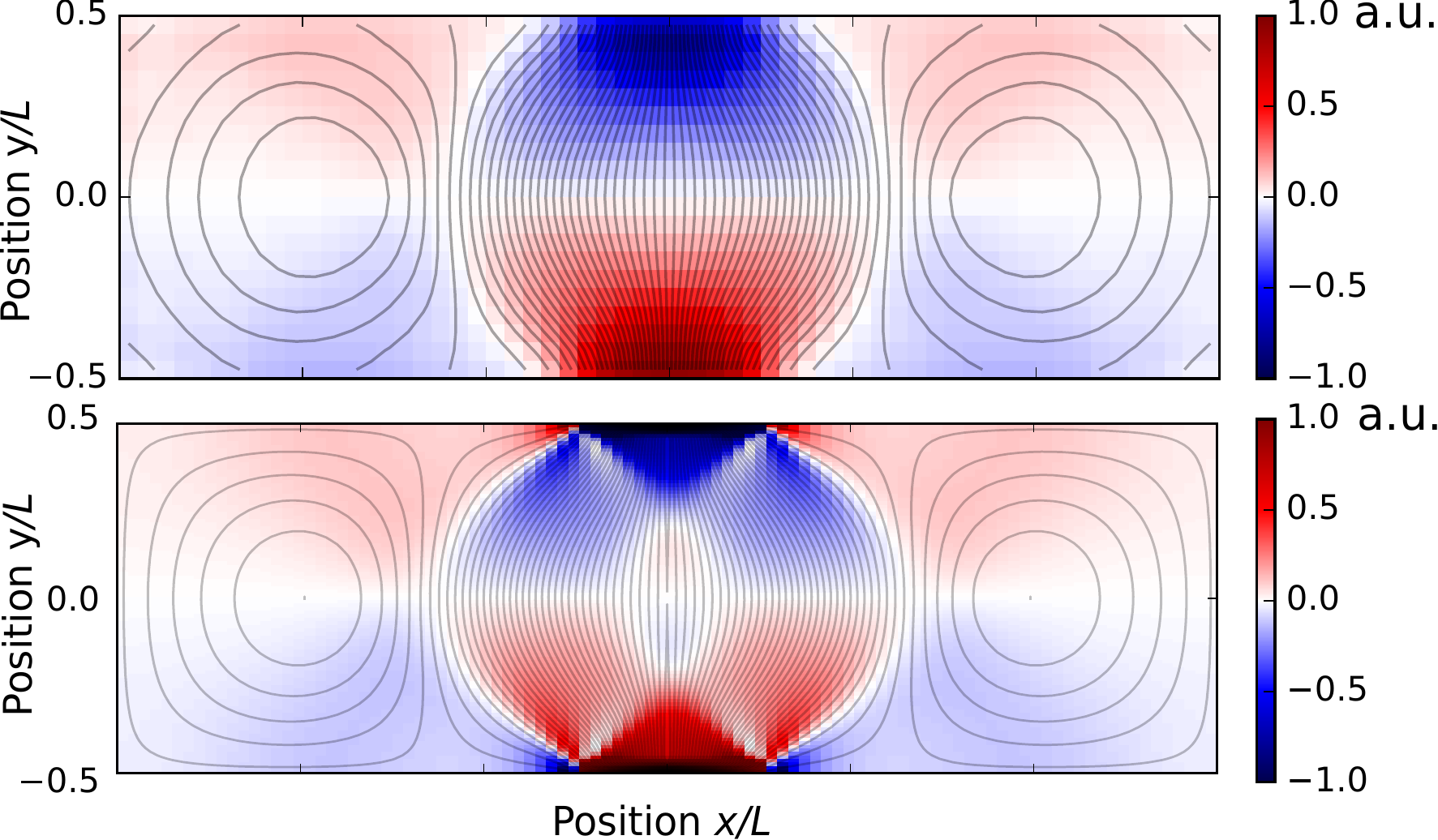}
\caption{\textbf{Comparison between numerical result and analytical approximation}.  Top panel: numerical simulation for the system for 60$\times$20 sites for $\gamma L/v=2$ (similar to the top panel in Fig. 2d in the main text). Lower panel: corresponding analytical solution represented by Eq.$\,$\eqref{eq:analyt_sol}.}
\label{figs:numerics_vs_analytical}
\end{figure*}

In Fourier space, the problem Eq.$\,$(4) and Eq.$\,$(9) in the main text with this source has a form in Fourier space
\be
\mathcal L \vec \Psi = \vec S,\qquad
\mathcal L= -\left( \begin{matrix}
 0 & iq_x & iq_y \\
2iv^2q_x &\gamma(1 +\lambda^2(q_x^2+2q_y^2)) & 2\gamma\lambda^2q_xq_y \\
2iv^2q_y & 2\gamma\lambda^2q_xq_y& \gamma(1 +\lambda^2(q_y^2+2q_x^2))
\end{matrix}\right)
\ee
where $\Psi = (P,J_x,J_y)$ as in the previous section, and the wavevectors are discrete, $q_x = 2\pi n/L$, $q_y = \pi m/L$ for integers $n$ and $m$. As in the previous section, $\lambda = v/\gamma$ has dimension of length and plays a role of mean free path. The solution can be presented in the form $\vec \Psi =  \mathcal L^{-1}\vec S$, where the inverse matrix is
\be 
\mathcal L^{-1} =\frac 1 {Q^2 }\left(
\begin{array}{ccc}
C_{\vec q}/2v^2& iq_x\Gamma_{x}/2v^2 & iq_y\Gamma_{y}/2v^2 \\
iq_x\Gamma_{x} & q_y^2 & q_yq_x \\
iq_y\Gamma_{y} & q_xq_y & q_x^2  \\
\end{array}
\right),\quad 
\ee
where we used the notations
\be
\begin{split}
&\Gamma_{x} = 1+\lambda^2(2q_x^2-q_y^2),\qquad \Gamma_{y} = 1+\lambda^2(2q_y^2-q_x^2)
\\
&Q^2 = q_x^2+q_y^2+2\lambda^22\lambda^4(q_x^4+q_y^4-q_x^2q_y^2),\quad C_{\vec q} = 1+3\lambda^2(q_x^2+q_y^2)+\lambda^4(2q_x^4+2q_y^4+q_x^2q_y^2)
\end{split}
\ee
Following from this result, the density and current density components can be found as an inverse discrete Fourier transform,
\be
\begin{split}\label{eq:analyt_sol}
&P(x,y) = \frac {\gamma}{2v^2}\sum_{q_x,q_y}\frac{C_{\vec q}}{Q^2}\Phi(q_x,q_y)e^{-iq_xx-iq_yy}\\
&J_x(x,y) = \sum_{q_x,q_y}\frac{iq_x\Gamma_{q_x}}{Q^2}\Phi(q_x,q_y)e^{-iq_xx-iq_yy}\\
&J_y(x,y) = \sum_{q_x,q_y}\frac{iq_y\Gamma_{q_y}}{Q^2}\Phi(q_x,q_y)e^{-iq_xx-iq_yy}
\end{split}
\ee
where $\Phi(q_x,q_y)$ is the Fourier transformation of the source term,
\be
\Phi(q_x,q_y) = \frac{iI_0}{L_xL_y}\sin(q_yL_y/2)\frac{\sin(q_xw/2)}{q_xw/2}
\ee
The plots for particle density and current components are shown on Fig. \ref{figs:numerics_vs_analytical} (lower panel).\\


\textbf{Part 5: Numerical Simulation Methods}\\

To derive the exact solutions shown in Fig. 2 in the main text, we  use numerical simulations for a 2D system of finite size. For a system of physical size $L_x\times L_y$ we consider a lattice with dimensions $N_x\times N_y$, where $N_x = L_x R_x$ and $N_y = L_y R_y$. Here, $R_x$ and $R_y$ are space resolutions which can be varied depending on the accessible computational resources. At the moment of publications, the power of personal computer is enough to perform computations for $R_k\sim 10-100$. Then, the density operator $\rho$ is represented by a normalized matrix $N_xN_y\times N_xN_y$.
The evolution of the density matrix was simulated using the discrete time recurrence relation
\be
\rho_{n+1} = U^{}_{\Delta t}\rho_n U^{\dag}_{\Delta t}-\gamma L(\rho_n)\Delta t+S\Delta t+O(\Delta t^2),
\ee
where $\rho_n$ is the density matrix at the turn  $n$ corresponding to time $t_n=n\Delta t$, $\Delta t\ll{\rm min}(N^{-1}_x,N^{-1}_y,\gamma^{-1})$ is a time step, $L$ is a linear superoperator acting on the density matrix as
\be 
L(\rho)_{nn'} =
\begin{cases}
 0,\quad &n= n'\\
 \rho_{nn'},\quad &n\neq n'
 \end{cases}
\ee
and $S$ is a source term matrix defined separately for each problem (see below). The expression uses the infinitizemal unitary
\be
U_{\Delta t} = \exp(-iH_0\Delta t)
\ee
instead of the commutator in Eq.(3) in the main text to avoid the effects of the $O(\Delta t^n)$ corrections leading to the instability at longer times. The 2D Hamiltonian has a represntation in form of the matrix
\be
H_0 = R_xT_x\otimes I_y+R_yI_x\otimes T_y,\qquad 
I_k = 
\left(
\begin{array}{ccccc}
 1 & 0 & 0 & 0 & .. \\
 0 & 1 & 0 & 0 & .. \\
 0 & 0 & 1 & 0 & .. \\
 0 & 0 & 0 & 1 & .. \\
 .. & .. & .. & .. & .. \\
\end{array}
\right)
,\qquad T_k = 
\left(
\begin{array}{ccccc}
 0 & 1 & 0 & 0 & .. \\
 1 & 0 & 1 & 0 & .. \\
 0 & 1 & 0 & 1 & .. \\
 0 & 0 & 1 & 0 & .. \\
 .. & .. & .. & .. & .. \\
\end{array}
\right)
\ee
where $I_x$,$T_x$ and $I_y$, $T_y$ are $N_x\times N_x$ and $N_y\times N_y$ matrices correspondingly, and $A\otimes B$ denotes the tensor product of matrices $A$ and $B$.

The initial value for the density matrix $\rho_0$ can be chosen arbitrarily since steady state for $\gamma\neq0$ does not depend on the initial conditions. For a particular system geometry, good choice of initial conditions may facilitate the fast convergence of the density matrix to the steady state. Nevertheless,  the density matrix can be chosen as identity matrix, $\rho_0 = \mathcal N^{-1} I_x\otimes I_y$, without a loss of generality, where $\mathcal N = N_x N_y$.

We assume that the steady state $\rho_\infty$ exists and can be approximated by matrix $\rho_n$ at time step $n$ with an error $\ve$ if
\be
||\rho_{n+1}-\rho_n||_1 \leq\ve
\ee
For normed density matrix $||\rho||\sim 1$ therefore, using the triangle inequality, one may suggest that reliable result is achieved if $\ve\ll 1$. In practice, for many cases $\ve$ can be as small as $\ve = 10^{-1}-10^{-2}$. Importantly, the number of steps required to achieve the particular error $\ve$ increases rapidly with $\ve^{-1}$ and $\gamma^{-1}$.

The the particle density and the current density can be derived using the matrix elements of the density operator,
\be
P(x,y,n\Delta t) = \frac N{L_xL_y}\rho_n(m,m), \quad J_x(x,y,n\Delta t) = \frac{NvR_x}{L_xL_y} \Im \rho(m,m+N_y), \quad J_y(x,y,n\Delta t) = \frac{NvR_y}{L_xL_y} \Im \rho(m,m+1)
\ee
where $m = (N_y[N_x x/L_x]+[N_yy/L_y])$, $[x]$ denotes the floor function, and the coefficients depend on the total number of particles in the system $N$ and the hopping parameter $v$. Note, that $m$ varies here from $0$ to $N_xN_y-1$ as in many programming languages such as Python and C++.

To visualize the current stream lines, one can use the stream function,
\be
\psi(x,y) = \int_C d\vec r \times \vec J,
\ee
where $C$ is an arbitrary trajectory between points with coordinates $x=0$, $y=0$ and $x$, $y$. The contour lines for the function $\psi(x,y)$  are shown in Fig. 2d in the main text and Fig.$\,$\ref{figs:numerics_vs_analytical} as the stream lines.
\\

\textbf{Flow between two surface planes}.
Exact vanishing the current at the boundary is a hard problem from numerical prospective described above since it requires arbitrarily small error $\ve$. Therefore it is more convenient to use the linearity of the problem, as we show in the previous sections, and present the solution in the form of superposition of two different solutions $\rho = \rho_1+\rho_2$,
where we assume $\rho_1$ to be a solution of homogeneous problem with a current density $J_x=0$, $J_y = J_0$ at any point of space. The density operator $\rho_2$ can be found as a solution for a problem with current source $S$ acting at the boundaries representing collisions at the surfaces. Numerically, this effect can be simulated using the source function
\be
 S = -i\gamma_s P\otimes T^{c}_y, \qquad P = \left(
\begin{array}{ccccc}
 1 & 0 & .. & 0 & 0 \\
 0 & 0 & .. & 0 & 0 \\
 .. &.. & .. & .. & .. \\
 0 & 0 & .. & 0 & 0\\
 0 & 0 & .. & 0 & 1 \\
\end{array}
\right),\qquad 
T^{c}_y = 
\left(
\begin{array}{ccccc}
 0 & 1 & 0 & 0 & .. \\
 -1 & 0 & 1 & 0 & .. \\
 0 & -1 & 0 & 1 & .. \\
 0 & 0 & -1 & 0 & .. \\
 .. & .. & .. & .. & .. \\
\end{array}
\right)
\ee
where $\gamma_s$ is chosen such that current vanishes at the boundaries, $J_y(\pm L_x/2,y) +J_0 = 0$. In practice, we can derive the solution of numerical problem with arbitrarily chosen $\gamma_s$ (e.g. $\gamma_s=1$) first, and then combine it with an appropriate current density value $J_0$. Since we are interested in normalized current density or the resistance, the value of $J_0$ drops from the final result.\\

\textbf{Two contacts configuration}. This problem involves the source and the drain of particles described by the source term
\be
S = I_s W \otimes P',
\qquad W =\left(
\begin{array}{ccc}
 0 & 0 & 0  \\
 0 & I_w & 0  \\
0 &0 & 0  \\
\end{array}
\right)
\qquad P' = \left(
\begin{array}{ccccc}
 1 & 0 & .. & 0 & 0 \\
 0 & 0 & .. & 0 & 0 \\
 .. &.. & .. & .. & .. \\
 0 & 0 & .. & 0 & -1\\

\end{array}
\right)
\ee
where $W$ is block-structured matrix of size $N_x\times N_x$, $I_w$ is unit $[wR_x]\times[wR_x]$ matrix, zeros denotes all other zero entries in the matrix $W$.
%

\end{document}